\documentstyle[twocolumn,prbbib,aps,epsfig]{revtex} 

\begin{document}
\twocolumn[\hsize\textwidth\columnwidth\hsize\csname@twocolumnfalse\endcsname

\title{Disproportionation Phenomena on Free and Strained Sn/Ge(111)
and Sn/Si(111) Surfaces}
\author{G.~Ballabio$^{1,*}$, G.~Profeta$^{2}$, S.~de~Gironcoli$^{1}$,
        S.~Scandolo$^{1,3}$, G.E.~Santoro$^{1}$, and
        E.~Tosatti$^{1,4}$}
\date{\today}

\address{%
$^{1}$International School for Advanced Studies (SISSA) and\\
Istituto Nazionale di Fisica della Materia (INFM), Unit\`a SISSA,
via Beirut 2-4, I-34014, Trieste, Italy.\\
$^{2}$Istituto Nazionale di Fisica della Materia (INFM), Unit\`a
dell'Aquila, and\\
Physics Department, Universit\`a dell'Aquila,
I-67010 Coppito (L'Aquila), Italy\\
$^{3}$Princeton Materials Institute, Princeton University,
Princeton, NJ 08540\\
$^{4}$International Centre for Theoretical Physics (ICTP),
P.O. Box 586, I-34014, Trieste, Italy.\\
}

\maketitle

\begin{abstract}
Distortions of the $\sqrt3\times\sqrt3$ Sn/Ge(111) and Sn/Si(111)
surfaces are shown to reflect a disproportionation of an integer
pseudocharge, $Q$, related to the surface band occupancy.
A novel understanding of the $(3\times3)$-1U (``1 up, 2 down'') and 2U
(``2 up, 1 down'') distortions of Sn/Ge(111) is obtained by a
theoretical study of the phase diagram under strain. 
Positive strain keeps the unstrained value $Q=3$ but removes
distorsions.
Negative strain attracts pseudocharge from the valence band causing
first a $(3\times3)$-2U distortion ($Q=4$) on both Sn/Ge and Sn/Si,
and eventually a $(\sqrt3\times\sqrt3)$-3U (``all up'') state with
$Q=6$.
The possibility of a fluctuating phase in unstrained Sn/Si(111) is
discussed.
\end{abstract}


]

Disproportionation, namely the spontaneous differentiation of
initially equivalent sites into states of unlike valence or charge, is
well known in solid state chemistry, for example in organic
charge-transfer salts\cite{nad} and also in transition metal and mixed
valence rare earth compounds\cite{falicov}.
Common features are an enlarged unit cell, due to ordering of the
disproportionated sites, and the presence of a well defined quantum
number (e.g., the valence) that is integer, even when the actual
charge inbalance is generally not integer.
Here we propose that a disproportionation, driven in this case by a
different physics, but equally characterized by integer valencies, is
at the origin of the well known but only incompletely understood
two-dimensional phase transition\cite{carpinelli} observed in 1/3
monolayer Sn-covered Ge(111) and Si(111) surfaces.
Moreover we present theoretical suggestions that the disproportionated
state could be changed, or even removed altogether, by external
strain.

The 1/3 monolayer Sn-covered Ge(111) and Si(111) surfaces have
received considerable attention since the discovery of a transition
from a room-temperature phase with $\sqrt3\times\sqrt3 R30^\circ$
periodicity to a low-temperature $3\times3$ phase, first seen in
Pb/Ge(111), then in Sn/Ge(111)\cite{carpinelli,goldoni}, and
unsuccessfully searched for in Sn/Si(111)\cite{ottaviano}.
The main reason of interest lies in the electronic driving force
behind the transition, namely the presence of a half filled surface
state, suggestive\cite{carpinelli} of a two-dimensional surface
charge-density wave (SCDW) mechanism\cite{tosatti}.
Recent X-ray data show that the transition is accompanied by static
freezing and long-range ordering of a $3\times3$ periodic structural
distortion, consisting mainly of a ``buckling'' of Sn adatoms by
$\sim0.35$~\AA{} along the surface normal\cite{zhang,bunk}.
In the distortion, adatoms are generally believed to follow a ``1 up,
2 down'' (1U) pattern, but there is evidence that under some as yet
unclear circumstances a non-negligible portion of the surface distorts
with the opposite phase, namely a ``2 up, 1 down'' (2U)
pattern\cite{melechko,avila1}.
The precise mechanism underlying the phase transition has led to
debate.
While the driving role of the partly filled surface states is
unquestioned, the 2D Fermi surface area and nesting properties of
Sn/Ge(111) do not fit a $3\times3$ periodicity as expected for
SCDWs\cite{santoro,scandolo}.
Due to the large electron-electron repulsion, a Mott insulator or a
spin-density wave state\cite{santoro} with negligible distortion could
prevail, as in the isoelectronic case of SiC(0001)\cite{anisimov}.
The large observed distortion and the order-disorder nature of the
phase transition\cite{avila2} however rule out both the weak-coupling
SCDW and the Mott insulator.
First-principles electronic structure calculations, initially
negative\cite{carpinelli,scandolo}, later correctly favored the 1U
distortion\cite{ortega,degironc}, with surface electronic bands in
agreement with experiment\cite{goldoni,avila2,uhrberg}.
The calculated energy gain was about 10 meV/adatom\cite{ortega},
mostly of Jahn-Teller nature.
Multi-band and bond-density wave aspects were also pointed out as
crucial factors that stabilize the 1U distortion relative to the Mott
insulator\cite{degironc}.
Surprisingly, these calculations showed no energy gain for a 2U
distortion\cite{degironc}, contrary to the known readiness of 2U to
show up experimentally in Sn/Ge\cite{melechko,avila1}.
The order-disorder character of the phase transition, well confirmed
by simulations\cite{avila2}, is in agreement with the persistence of a
1U-like Sn $4d$ core level splitting in nominally undistorted Sn/Ge at
high temperature\cite{mascaraque}.
However a puzzling 2U-like persistent core splitting in apparently
undistorted Sn/Si\cite{uhrberg,lelay} remains unexplained.
Finally, substitutional defects play an important role.
A Ge (Si) adatom replacing Sn acts as an electron donor in the Sn/Ge
(Sn/Si) surface, triggering a local distortion of roughly 2U
type\cite{melechko}.
However the defect is also likely to exert a negative local strain,
effectively squeezing the neighboring Sn adatoms outwards relative to
the substrate, so that strain and charge donation apparently conspire.
One is eventually led to wonder about the general role of strain, and
about whether some of these open questions could be addressed by
studying primarily the effect of strain on these surfaces.
We report here the surprising results of a theoretical study of the
Sn/Ge(111) and Sn/Si(111) systems under uniform strain.
Besides predicting new strain-driven phase transitions, we find that
these surface distortions are in the nature of disproportionation
phenomena.

We carried out electronic structure calculations within the gradient
corrected\cite{perdew} local density approximation (GC-LDA) where, 
in the absence of strain, a stable ($3\times3$)-1U distortion was
established for the Sn/Ge(111) surface\cite{ortega,degironc}, whereas
Sn/Si(111) is still marginally undistorted\cite{perez}.
We used slab geometries consisting of three Ge (Si) bilayers, a planar
$3\times3$ cell with 18 atoms/bilayer, 3 Sn adatoms per cell in the
$T_4$ position, and 9 hydrogen atoms saturating the bottom surface of
the slab.
We employed the PWSCF parallel code\cite{pwscf.org} with a well
checked plane-wave cutoff of 12 Ry\cite{ballabio1}, and k-summation
schemes based on 6 Monkhorst-Pack k-points in the irreducible zone,
extended when necessary to 16 and 36 k-points.
We strained the slab cell isotropically in the surface plane, by
increasing or decreasing its $(x,y)$ sides by up to $\pm5\%$, full
relaxation allowed along $z$, the surface normal, so that inside the
slab the three dimensional strain tensor was effectively uniaxial.
For each strain value, the stable surface geometry was sought for
through a series of relaxation runs performed by constraining the
value of the adatom buckling $\Delta z$, (the $z$-difference between
``up'' and ``down''adatoms) and relaxing all other atomic coordinates,
except for the lowest Ge (Si) monolayer, held fixed and bulk-like (or
strained), along with the saturating H layer.

\begin{figure}
 \centerline{\epsfig{file=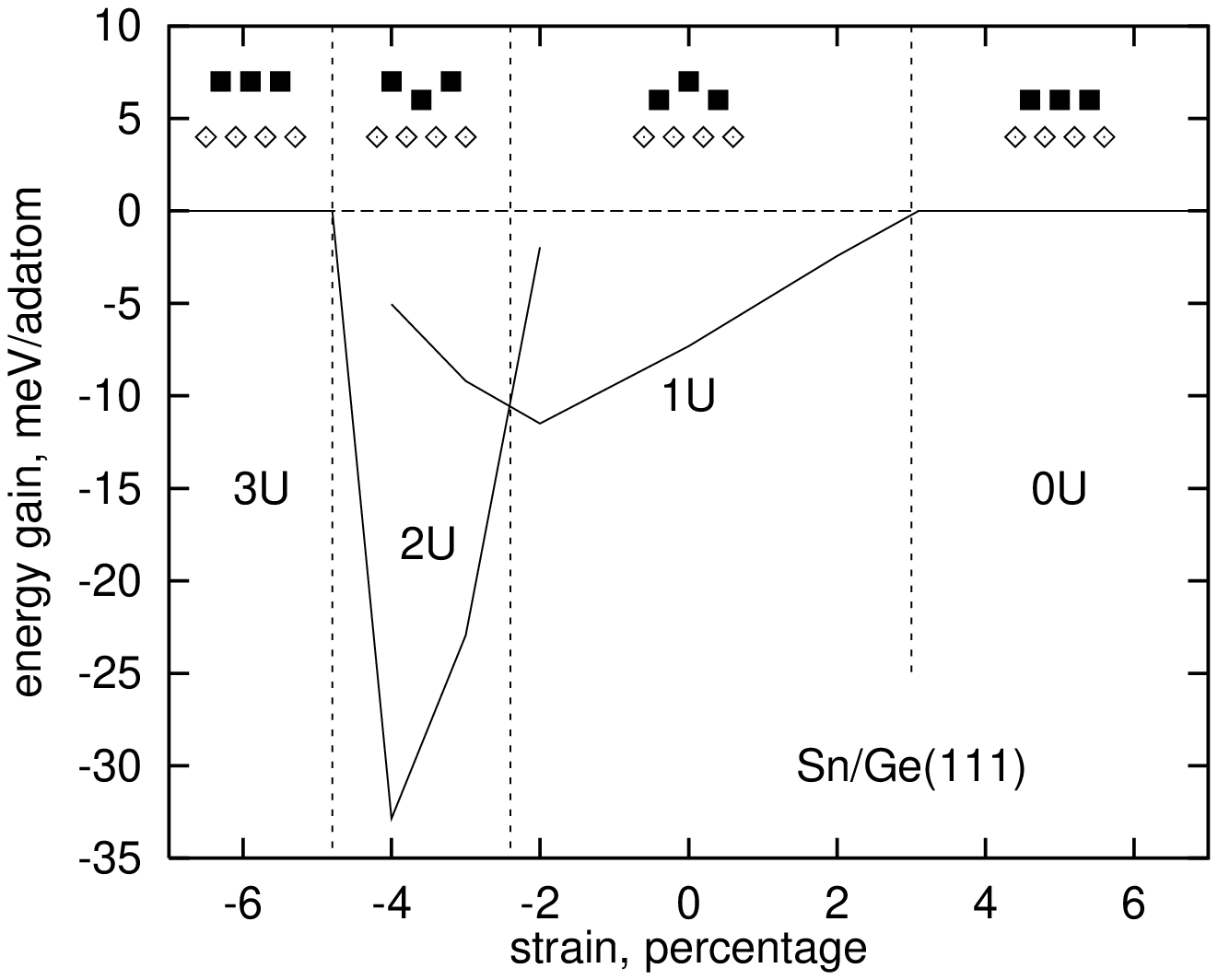,angle=270,width=0.48\textwidth}}
 \centerline{\epsfig{file=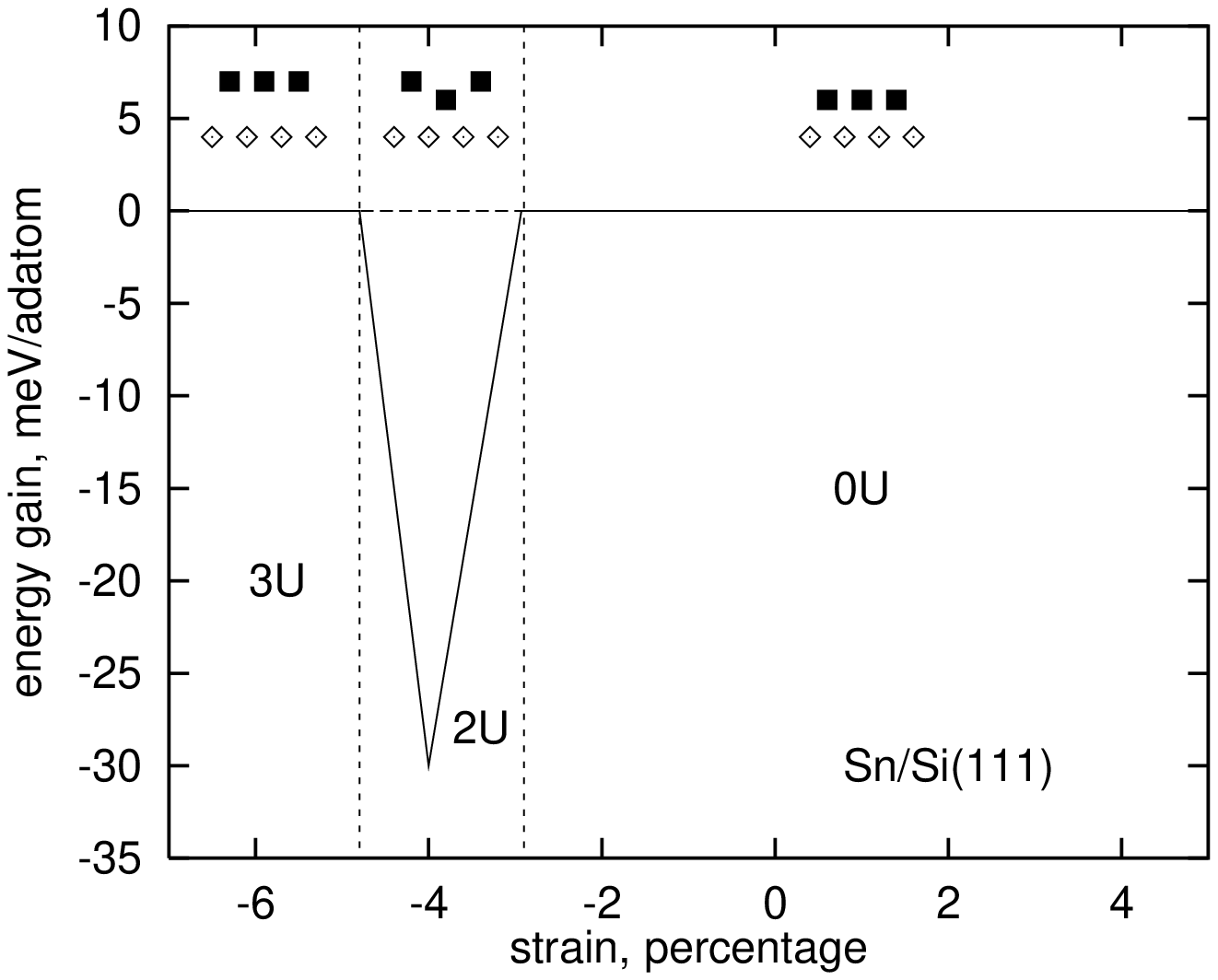,angle=270,width=0.48\textwidth}}
 \caption{Energy gained by the $(3\times3)$-1U and 2U distortions
against the relaxed undistorted $\protect\sqrt3\times\protect\sqrt3$
phase, as a function of strain.
(a) Sn/Ge(111);
(b) Sn/Si(111).}
 \label{fig:energygain}
\end{figure}

Figure \ref{fig:energygain} summarizes the zero-temperature phase
diagram versus strain, the order parameter and energy gain being given
for the $3\times3$ distorted state relative to the corresponding fully
relaxed $\sqrt3\times\sqrt3$ state at the same strain, for both Sn/Ge
and Sn/Si.
At zero strain, in agreement with experiment and with previous
calculations\cite{ortega,degironc,perez}, Sn/Ge is 1U distorted, with
$\Delta z \sim 0.34$~\AA{}, whereas Sn/Si is 0U (undistorted).
Positive strain (in-plane expansion) causes Sn/Ge to lose its
distortion, through a 1U-0U transition.
By mechanically attracting the Sn adatoms closer to the Ge substrate,
strain pushes the half filled dangling bond band upwards in energy
towards the middle of the gap.
In line with our earlier finding that the $3\times3$ distortion is a
multi-band effect\cite{degironc}, in this restored single-band
situation the surface reverts to the undistorted $\sqrt3\times\sqrt3$
state.
The resulting narrow half-filled surface band metallic state should
actually be prone to other instabilities.
We did check with spin-polarized calculations for a possible
antiferromagnetic (Mott-Hubbard) state similar to that of
isoelectronic SiC(0001); but up to 5\% strain we did not find one, at
least in Sn/Ge.
Negative strain (in-plane compression) on the contrary pushes the
adatoms upwards and that, beyond a critical strain of $\sim-3\%$,
stabilizes a 2U distortion, with a 1U-2U transition in Sn/Ge, and a
0U-2U transition in Sn/Si. 
Within our approximation, the 2U order parameter $\Delta z$ at
$\varepsilon=-4\%$ is 0.49~\AA{} with an energy gain with respect to
the 0U configuration of 39.5 meV/adatom in Sn/Ge, and 0.4 \AA{} with
an energy gain of 30 meV/adatom in Sn/Si.
In both cases the energy gain is suggestive of a 2U-disordered
transition at relatively large critical temperature.
At still larger negative strains $\sim-5\%$, all adatoms are finally
forced up, with a 2U-3U transition, and an optimal 3U state for both
Sn/Ge, and Sn/Si.
This state should be seen as uniformly distorted -- same symmetry as
the ``0U'' undistorted $\sqrt3\times\sqrt3$ state, but with quite
different physics as we shall see.

\begin{figure}
 \centerline{\epsfig{file=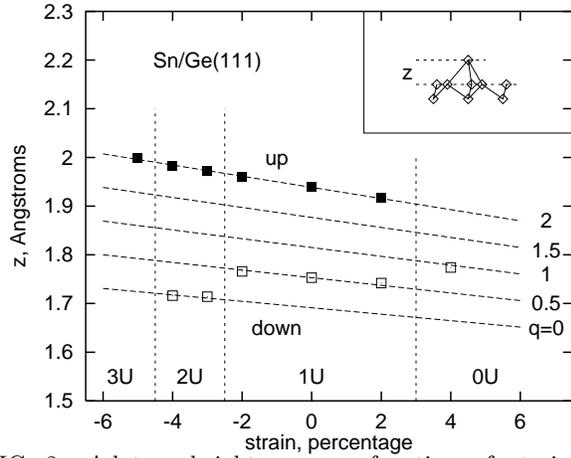,angle=270,width=0.48\textwidth}}
 \caption{Adatom height $z$ as a function of strain in Sn/Ge(111).
Open squares: ``down'' adatoms;
filled squares: ``up'' adatoms;
dashed lines: optimal height of an isolated charged adatom (see
text).}
 \label{fig:quantization}
\end{figure}

Closer examination of the distorted geometries reveals a striking
feature: the Sn adatoms' height above the substrate is bimodal,
approximately quantized for all strains and states.
A given Sn adatom can be either U or D, but not in between.
This is reminiscent of static mixed valence, or disproportionated
states, where an individual ion site is endowed with a choice between
two integer valence states, the overall proportion between them
controlled by a chemical potential.
The valence state can either fluctuate dynamically with time, or be
frozen in a static regular or irregular spatial pattern.
In that case, all local properties, including the surrounding ion
geometry, display a bimodal distribution.
We found, by analysing the electronic structure, that in our case the
role of effective valence is played by a {\em surface state occupancy}
$q$, which is 2 in all U adatoms, 1 in all adatoms of the undistorted
(``0U'') phase, 0.5 in D adatoms of the 1U phase of Sn/Ge, and zero in
the D adatoms of 2U phases.
This conclusion was established with the help of a separate Sn/Ge
calculation, where a single Sn adatom was confined in the $3\times3$
slab cell.
The six remaining first-layer Ge atoms were saturated by six
hydrogens, effectively decoupling the Sn adatom from its replicas.
When a variable electron number was introduced with a net electron
excess between $-1$ and $+1$ (charge neutrality obtained through a
matching uniform positive background), it was found that the
population change was restricted to the Sn dangling bond surface
state, its occupancy $q$ correspondingly varying from zero to 2.
For each $q$ value, the total energy was minimized as a function of
adatom height $z$ over the substrate, and the optimal
$z_0(q)$ was determined.
The calculations were repeated for various strains and the results
reported in Fig. \ref{fig:quantization} (dashed lines).
The comparison of the calculated $z_0(q)$'s with the individual
adatom heigths previuosly calculated in the 0U, 1U, 2U and 3U stable
phases of Sn/Ge is striking and provides a strong evidence of
quantization of the individual $q_i$'s at four and only four heigth
values, corresponding within a few percent error to the occupancies 2,
1, 1/2, and 0 indicated above.
These quantized charges $q_i$ are actually pseudocharges, not to be
confused with real, physical electron charges on the adatoms.
An approximate Mulliken analysis of valence charges\cite{ballabio2}
showed in fact that the actual charge redistribution among adatoms is
always less than 0.1 $e$. 
A negligible redistribution of true, total charge is also in good
accord with the observed $4d$ Sn core level splitting, consisting of
one deep and one shallow levels in a 1:2 intensity proportion,
attributable to U and D adatoms respectively.
Had the true charge inbalance even only resembled that of the
pseudocharges, $2$ and $0.5$ for U and D adatoms, respectively, the
sign of the splitting should have been the opposite, and its magnitude
much larger.
We calculated\cite{ballabio2} the magnitude of the core level
splitting between U and D adatoms in Sn/Ge-1U in the
final-state\cite{pehlke} approximation, and obtained 0.20~eV, in
reasonable agreement with 0.32~eV, observed both in ($3\times3$)-1U
and for the high temperature $\sqrt3\times\sqrt3$ phase of
Sn/Ge(111)\cite{avila2,lelay,gothelid}.
The split core level intensity ratio should be instead 2:1 for a 2U
distorted surface, the U and D adatoms being now in that proportion,
opposite to the 1:2 of the 1U case.
Repeating our core level calculation for the $-3\%$ strained
Sn/Si(111) ($3\times3$)-2U surface we obtained a Sn $4d$ splitting of
$-0.20$ eV, in qualitative agreement with $\sim-0.4$ eV observed on
the nominally {\em undistorted} Sn/Si(111) $\sqrt3\times\sqrt3$
surface.
We shall return to this point below.

\begin{figure}
 \centerline{\epsfig{file=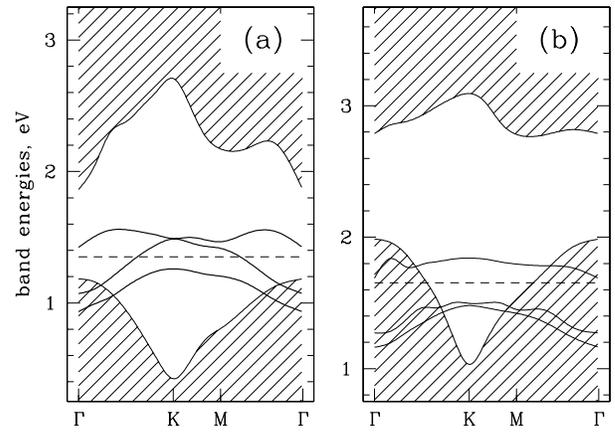,width=0.48\textwidth}}
 \caption{Electron bands of Sn/Ge(111) $3\times3$.
(a) 1U, zero strain;
(b) 2U, strain $-$4\%.}
 \label{fig:bands}
\end{figure}

We are now in a position to address directly the physics behind the
pseudocharge and its disproportionation, with and without strain, in
connection with the electronic structure.
In the 0U $\sqrt3\times\sqrt3$ phases, the surface state band is
isolated inside the gap, each adatom contributing one electron to it,
whence $q_i=1$ and $Q=\Sigma q_i=3$.
This is an undistorted metal state, at least within GC-LDA.
In the ($3\times3$)-1U phase, the surface band hybridizes much more
with the valence bands, especially with the Ge-Ge bonding state
directly underneath the adatom\cite{degironc}.
There is still on average one electron/adatom; however the total
pseudocharge $Q=3$ per cell disproportionates, $q=2$ electrons now
filling the U adatom's dangling bond (Figure \ref{fig:bands}a), the
remaining one shared between the other two D adatoms, each with
$q=0.5$.
The onset of the ($3\times3$)-1U ground state in Sn/Ge is thus favored
by the small Ge bandgap, and by the consequent energy proximity of the
surface band to bulk valence bands.
That is consistent with the absence of distortion in Sn/Si(111), and
in SiC(0001), where the surface band is more isolated in the wider
bandgap.
In the strain-induced 2U phase, both U adatoms have $q=2$, while D has
$q=0$.
The total pseudocharge $Q$ is now 4: one extra electron has been drawn
from bulk valence bands into the surface band.
For strain values in the 2U range, the surface bands (Figure
\ref{fig:bands}b) drop in fact enough in energy to overlap the bulk
valence bands, now slightly above $E_F$.
Furthermore, because the difference $q_{U}-q_{D}$ is greater than in
the 1U phase (2 compared to 1.5), the optimal distortion is larger
(0.49 \AA{} compared to 0.34 \AA{} in Sn/Ge).
The 2U distortion amplitudes and energy gains are correspondingly
smaller in the stiffer, wide-gapped Sn/Si.
Finally, in the ($\sqrt3\times\sqrt3$)-3U phase, all adatoms are
equivalent with $q=2$.
The surface band lies entirely below $E_F$, and 3 electrons are
transferred from bulk to surface, so that $Q=6$.
In this sense, the 3U state is fundamentally different from the 0U
state, despite its sharing the same symmetry.
In all these distorted states, the surface state band breaks up into
narrow adatom-localized levels.
Their filling and emptying leads to a Jahn-Teller-like energy gain,
and that explains the quantization of the pseudocharge $q$.
Negative strain acts as a chemical potential, and disproportionation
arises out of the necessity to distribute an increasing electron
number through integer surface state occupancies.
In chemical terms, U adatoms could perhaps be compared with
``tetravalent'' Sn, D to ``divalent'' Sn\cite{waghmare}.

These results and predictions should be amenable to experimental test,
particularly if the strains required to provoke the transitions could
realistically be introduced, e.g., by heteroepitaxy.
Even if our calculated critical strain values are probably affected by
systematic errors of order 1--2\%, connected with lattice parameter
errors and with imperfect handling of incipient strong correlations by
the density functional calculation, the strain-induced sequence of
transitions seems inescapable.
Among the phenomena that are already observed, the 2U patches reported
on Sn/Ge\cite{melechko} seem now very likely to be induced by some
local strain.
The puzzling 2U core level signature of nominally unstrained Sn/Si is
likely to be due to a dynamical, fluctuating 2U state.
Assuming in fact a 0U-2U zero-temperature, classical transition to
occur in reality at some small positive strain (instead of a small
negative strain as in the GC-LDA calculation), the long range order
could simply be destroyed by thermal fluctuations, still leaving a
local 2U distorted structure\cite{modesti}.
Of additional interest is the possibility of {\em quantum mechanical}
fluctuating states, analogous to dynamical mixed
valence\cite{falicov}.
Close to the critical strain values separating two classical
distortion states, such as 0U-1U, 0U-2U, 1U-2U, 2U-3U, tunneling
across low barriers could lead to quantum melting of the classical
long-range order.
This kind of polaronic state could entail the exotic but real
possibility of {\em surface state superconductivity} when, as the case
for 0U-2U, the pseudocharge fluctuation takes place in units of two,
thus realizing a bipolaron state.
The large distortions involved suggest however that this novel
strong-coupling surface polaronic metals and bipolaronic
superconductors might be attainable only at very low temperatures.

This work was supported by MIUR through COFIN, and by INFM through INFM/F and 
"Iniziativa Trasversale Calcolo Parallelo".
We are grateful to S.~Modesti for his invaluable help.

\clearpage

\end{document}